\def\minone{$^{-1}$}
\def\sqiggt{\hbox{\rlap{\lower.55ex \hbox {$\sim$}}
\kern-.3em \raise.4ex \hbox{$>$}\,}}
\def\sqiglt{\hbox{\rlap{\lower.55ex \hbox {$\sim$}}
\kern-.3em \raise.4ex \hbox{$<$}\,}} 
\def\mbold#1{\mbox{\boldmath$#1$\unboldmath}}
\def\kev{\,ke\kern-.1em V} \def\ev{\,e\kern-.1em V} 
\def\sqig{$\sim\,$} \def\etal{et\,al.} 
\def\appro{$\approx\,$}\def\msun{M$_{\scriptstyle\odot}$} 
\def\up#1{$^{\mbox{{\scriptsize #1}}}$}  
 \def\pten#1{$\times10^{#1}$}
\def\deg{$^{\circ}$} \def\kmps{km\,s$^{-1}$} 

\def\vcen{V1025~Cen}
\def\beat{$\omega$\,--\,$\Omega$}

\def\Hb{H$\beta$}\def\Hg{H$\gamma$}
\def\HeI{He\,{\sc i}}\def\HeII{He\,{\sc ii}}

\def\Pspin{$P_{\rm spin}$}\def\Porb{$P_{\rm orb}$}
\def\rco{$R_{\rm co}$}\def\rcirc{$R_{\rm circ}$}
\documentstyle{mn}
\title[Accretion mode of V1025~Cen]
{On the accretion mode of the intermediate polar V1025~Centauri}
\author[C.~Hellier \etal]{Coel Hellier,\up{1}\ Graham A.\,Wynn\up{2}\ and
David A.\,H.~Buckley\up{3}\\
\up{1}Astrophysics Group, School of Chemistry and Physics, 
           Keele University, Keele, Staffordshire, ST5 5BG\\ 
\up{2}Astronomy Group, University of Leicester, Leicester LE1 7RH\\
\up{3}South African Astronomical Observatory, PO Box 9, Observatory 7935, 
Cape Town, South Africa}

\date{ }
\begin{document}
\maketitle
\begin{abstract}
The long white-dwarf spin periods in the 
magnetic cataclysmic variables EX~Hya and \vcen\ imply that if the
systems possess accretion discs then they cannot be in equilibrium. 
It has been suggested that instead they are
discless accretors in which the spin-up torques resulting from
accretion are balanced by the ejection of part of the accretion
flow back towards the secondary.   We present phase-resolved spectroscopy
of \vcen\ aimed at deducing the nature of the accretion flow, and compare
this with simulations of a discless accretor.  We find that both the
conventional disc-fed model and the discless-accretor model have 
strengths and weaknesses, and that further work is needed before 
we can decide which applies to \vcen.
\end{abstract}

 
\section{Introduction}
The magnetic cataclysmic variables are close binary stars 
in which one can study the interaction of an accretion flow with a
magnetic field. Where the accreting white dwarf is only weakly
magnetic (\sqiglt 10$^{5}$ G) an
accretion disc disc in a manner little different from that in non-magnetic
systems. Stronger fields (\sqiggt 10$^{7}$ G) lock the
spin of the white dwarf to the binary orbit and dominate the
accretion flow, forcing it to accrete along field lines. 
The intermediate case is less clear, and systems in this
category (refered to as intermediate polars or IPs) display a range of
behaviours depending on the mass-transfer rate,
field strength and white-dwarf spin period.  Among the 
possibilities are (1) a partial disc which
is disrupted when the magnetic pressure exceeds the ram pressure,
giving way to magnetically channelled flow inside the magnetosphere;
(2) a partial disc, but with some of the accretion stream overflowing
the disc to interact directly with the magnetosphere (e.g.\ Hellier
1991); (3) discless accretion in which the flow can be regarded as
diamagnetic, intermediate between the purely ballistic and magnetically 
channeled flows (e.g.\ King 1993; Wynn \& King
1995); (4) models in which the propeller effect of a rapidly spinning
field prevents accretion (e.g.\ Wynn, King \&\ Horne 1997).  Recent
reviews of these possibilities are presented in Hellier (2001), chapter 9,
and Wynn (2001).
  
Of particular relevance to this paper is the ratio of the spin period
of the white dwarf to the orbital period of the binary.  Most
IPs have \Pspin/\Porb\,\sqiglt 0.1, and indeed no
system can both possess an accretion disc and be in equilibrium unless
this inequality holds.  This condition is equivalent to the condition
\rco\,\sqiglt \rcirc\ where \rco\ is the corotation radius 
(the radius at which the magnetic field corotates with a
Keplerian flow) and \rcirc\ is the circularization radius (the radius
at which a circular orbit has the same angular momentum as the 
stream from the inner Lagrangian point). 
However, King \& Wynn (1999) discovered that a
discless system can reside on a continuum of equilibria with 
\rcirc\,\sqiglt \rco\,\sqiglt $b$, where $b$ is the  
distance to the Lagrangian
point. Such a system would have a longer spin period, with
0.1\,\sqiglt \Pspin/\Porb\,\sqiglt 0.7.  At the time only one IP 
(EX~Hya, with \Pspin\ = 67 mins and \Porb\ = 98 mins) was known to
have a secure \Pspin/\Porb\ ratio greater than \appro 0.1.  The
purpose of this paper is to (1) confirm earlier indications that
\vcen\ is a second system in this category, and (2) analyse
spectroscopic observations to investigate whether the accretion flow
is better described by the partial-disc model or by the
diamagnetic-flow model.

Buckley \etal\ (1998)'s discovery paper on \vcen\ (RX\,J1238--38) 
and follow-up X-ray observations (Hellier, Beardmore \&\ Buckley 1998)
found a spin
period of 2147 s (revealed by an optical and X-ray pulsation), and
suggested an orbital period near 85--90 mins, and thus a \Pspin/\Porb\ 
ratio of \appro 0.4.  The star is also notable for showing a 
1860-s optical and $J$-band periodicity
(Buckley \etal\ 1998). Given the above spin and orbital periods, the
only plausible identification is with the first harmonic of the beat
cycle between the orbital and spin cycles [i.e.\ the frequency
2(\beat) where $\omega$ and $\Omega$ are the spin and orbital
frequencies respectively]. Note, though, that no other IP shows a
lightcurve containing 2(\beat) but not \beat.  Other than this,  
\vcen\ is little studied, with, as yet, no
ephemerides for the periodicities, no estimate of the field strength,
and no determination of the binary inclination or of the component
masses. Note that a possible grazing eclipse reported 
by Allan \etal\ (1999) was an artefact of 
incorrect data reduction.

\section{Observations and data}
We observed \vcen\ with the 3.9-m AAT and the RGO spectrograph plus 
a TEK CCD. A 1200 lines 
mm\minone\ grating gave a resolution of 1.4\AA, covering the range 
\Hg\ to \Hb. Observing for 2 h, 4 h \&\ 3.5 h on the three consequtive
nights 1996 May 10--12 we obtained 320 integrations of 100 s each,
thus covering \sqig 7 orbital cycles and \sqig 16 spin cycle of this
star.  The summed spectrum, containing \HeI\ and \HeII\ lines in addition
to the Balmer lines, is shown in Fig.~1. It is interesting to note that
the spectrum is similar to that of EX~Hya (e.g.\ Hellier \etal\ 1987), in 
that the lines are broader than in most IPs.

\begin{figure} \vspace{7.6cm}      % Fig 1
\includegraphics{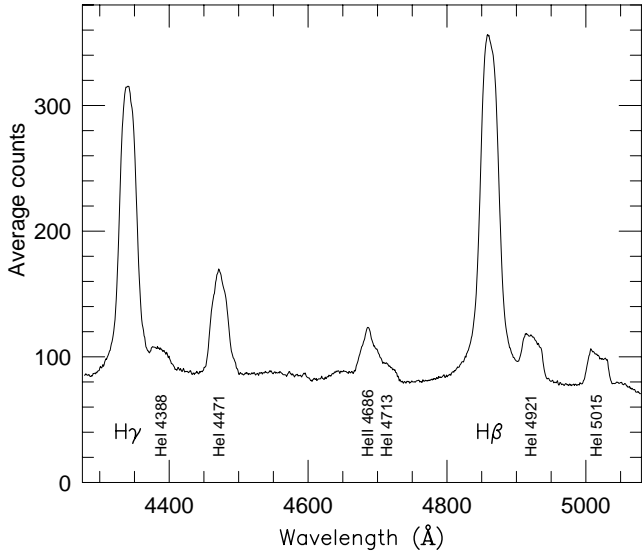}
\caption{The averaged spectrum of \vcen.}
\end{figure}

As a first look at the data we computed the equivalent widths 
and the V/R ratios for the lines in each spectrum (V/R being
the ratio of the equivalent widths on either side of the rest wavelength).
The Fourier transforms of these quantities for \Hb\ are shown in Fig.~2.

\begin{figure*} \vspace{7.0cm}      % Fig 2
\includegraphics{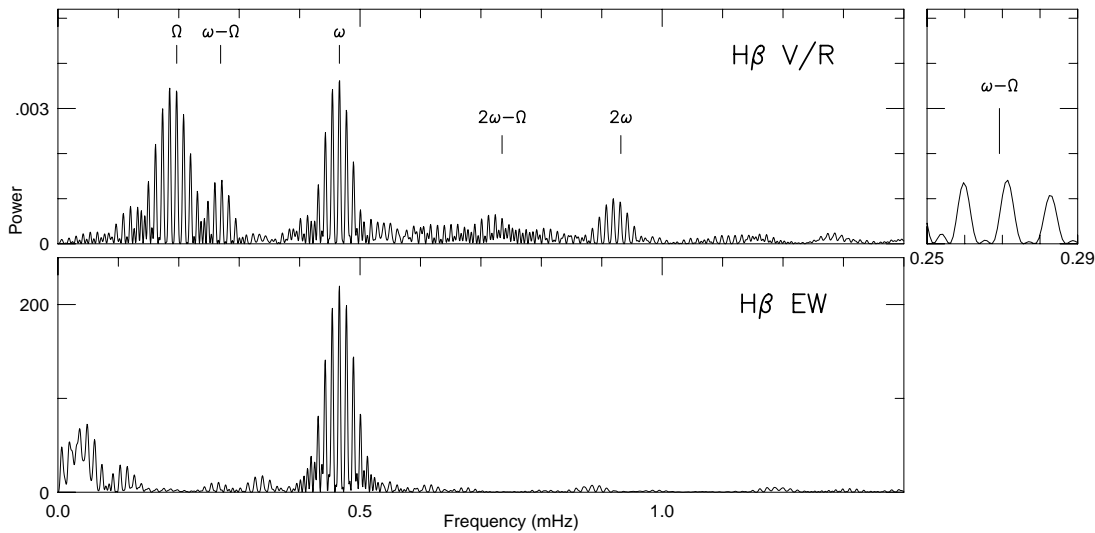}
\caption{Fourier transforms of the \Hb\ V/R ratios and equivalent widths.
The orbital frequency is labelled $\Omega$ and the spin frequency $\omega$.
At right the expanded section illustrates the frequency discrepancy of the
beat (\beat) pulse.}
\end{figure*}

Following Buckley \etal\ (1998) we identify the V/R periodicity 
near 0.2 mHz with the orbital cycle. From our data, though, we
cannot determine which of two 1-d aliases at 5090\,$\pm$\,30 and
5410\,$\pm$\,30 s (85 and 90 mins) is the true period.  However,
Buckley \etal\ used the proposed identification of the 1860-s photometric
period with 2(\beat) to derive (given the 2147-s spin cycle) an orbital
period of 5077 s, thus
favouring the 85-min alias. We therefore adopt 5077 s as the orbital
period, though its 1-d alias should be borne in mind.

The 2147-s spin period is prominent in both the V/R ratios and the
equivalent widths.  The V/R ratios also show power at 2$\omega$ and
possibly, though nearer the noise level, at 2$\omega$\,--\,$\Omega$.  A
peak near \beat\ is also clearly above the noise, but is shifted
significantly (by 0.7 per cent) from the expected frequency. We
discuss this later (Sections~5.2 \&\ 6.1).

\section{Orbitally resolved line profiles}
In Fig.~3 we show the line profiles of \Hb\ folded on the orbital
(5077-s) period. Each spectrum was first normalised to the continuum
level, so that the plot shows quasi equivalent widths (these are more
robust than fluxes in narrow-slit spectroscopy).  In constructing the
fold we are assuming that line-profile variations on the spin or other
cycles will smear out into a phase-invariant profile. Thus, to
emphasize the varying component, we additionally show the data after 
subtracting 
the phase-invariant profile. Also in Fig.~3 is the Doppler
tomogram of the subtracted profiles, computed using the
back-projection technique (see Marsh \&\ Horne 1988).  We present all plots
with phase 1 corresponding to a guess at when inferior conjunction of
the secondary occurs, but note that we have no secure knowledge of
this.

\begin{figure} \vspace{18cm}      % Fig 3
\includegraphics{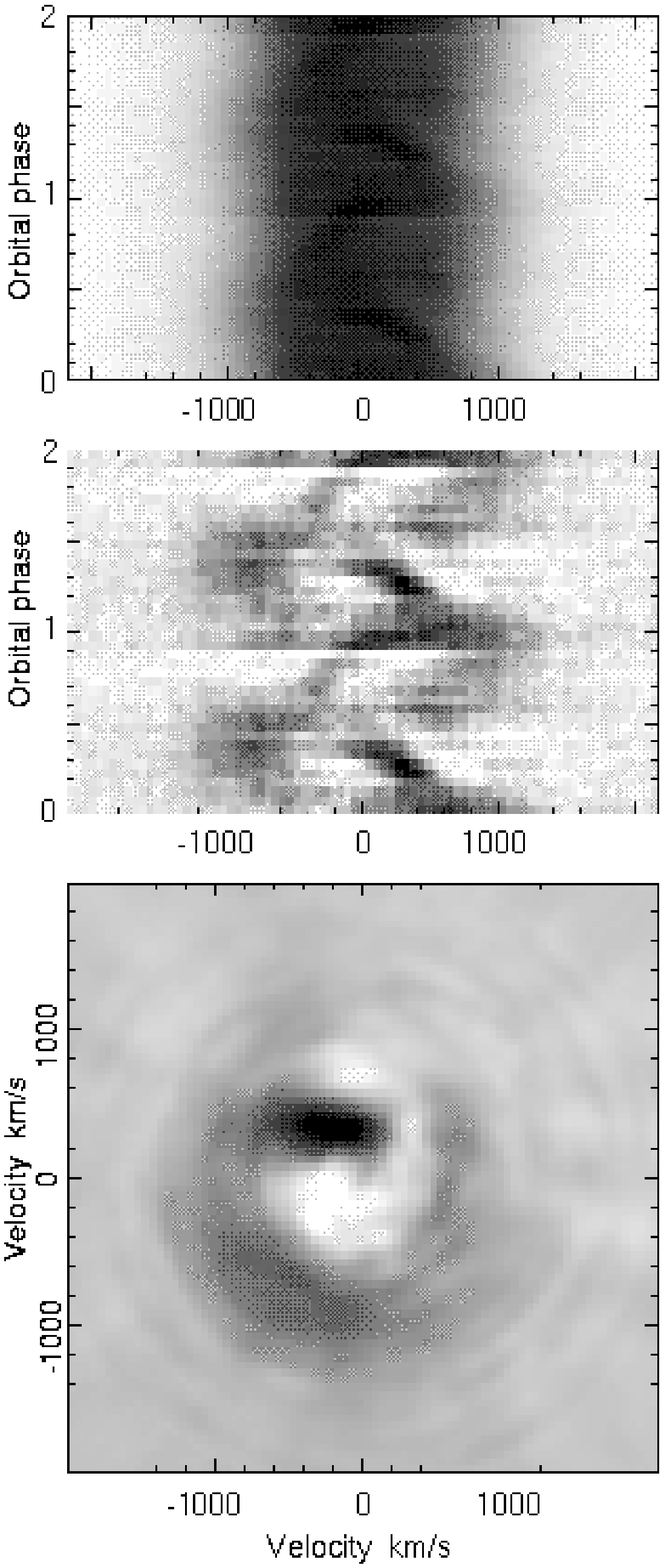}
\caption{The panels show the \Hb\ line folded on the orbital 
period (top), the same after subtraction of the phase-invariant profile,
and the corresponding Doppler tomogram (bottom).  The adopted phase zero 
is at HJD 2450213.87785.}
\end{figure}

In the line center is an S-wave with a projected velocity of \sqig 350
\kmps, phased with maximum redshift at \sqig 0.15 in our (insecure)
convention, and appearing as the brightest region of the tomogram.
Such as S-wave would conventionally be interpreted as arising from the
heated face of the secondary, or from the accretion stream, particularly
where it hits an accretion disc. 

There is also an ill-defined higher-velocity feature which has a
maximum blueshift near phase 0.4, when its velocity appears to be
centered at \appro 900 \kmps.
It is too ill-defined to allow us to measure the
amplitude of its orbital motion, but in the tomogram it gives
rise to a brightening in the lower-left quadrant.

There is weak evidence for a double-peaked structure, as would
arise from a disc. This would form a ring-structure in the tomogram,
centered on the velocity of the white dwarf.  It is possible to interpret
the tomogram in this way, though the ring is ill-defined and incomplete.

\section{Line profiles over spin phase}
Fig.~4 shows the variation of the line profiles over spin phase, again
displayed as quasi equivalent widths.
There is a prominent variation in equivalent width
with the whole line becoming brighter at phase 1 
(the phasing adopted is arbitrary, as we do not yet have an 
ephemeris to link it to the photometric or X-ray pulses).
Since the spin-cycle variation is primarily a change in equivalent
width, rather than velocity, it violates the constant-flux assumption
of tomography and the resulting tomogram is not useful [see 
Hellier (1999) for the tomogram and further discussion of this issue]. 

The spin-resolved line profiles look similar to those in EX~Hya (e.g.\ 
Hellier \etal\ 1987), and thus we adopt essentially the same 
interpretation (see also Buckley \etal\ 1998; Hellier 1999). In this model 
the spin-varying emission comes from the accretion curtains of
magnetically trapped material falling onto the magnetic poles. The
simultaneously bright red and blue wings and the general symmetry of
the profile then implies that we are seeing emission from both upper
and lower poles.  The brightest emission (phase 1) probably occurs
when the upper pole points away from the observer, and the view of the
white dwarf is unobscurred, allowing us to see the bright,
high-velocity regions of curtain near the white dwarf.  Half a cycle
later, the outer regions of the upper curtain are in front of the
white dwarf, obscurring the bright regions, resulting
in the fainter line seen at phase 0.5.

However, the lines are not symmetric, having a
brightening in the blue wing at phases 0.6--0.8 that is not seen in
the red wing. Thus the curtains of material must be asymmetric
or twisted; this might result from the fact that, as discussed
next, the field lines must 
rotate more slowly than the inner edge of any disc. 

\section{A conventional model?}
We first try to interpret the above results in the context of the
conventional model of an IP: one that accretes through a partial disc
which feeds field lines from its inner edge. Given the exceptionally long
spin period of \vcen, the system cannot be in equilibrium in this model,
and the field lines must rotate more slowly than
the Keplerian motion at the inner disc edge (if this were not the case
the magnetosphere would extend beyond the circularisation radius and
the angular momentum of the disc would dissipate).  

\begin{figure} \vspace{10.4cm}      % Fig 4
\includegraphics{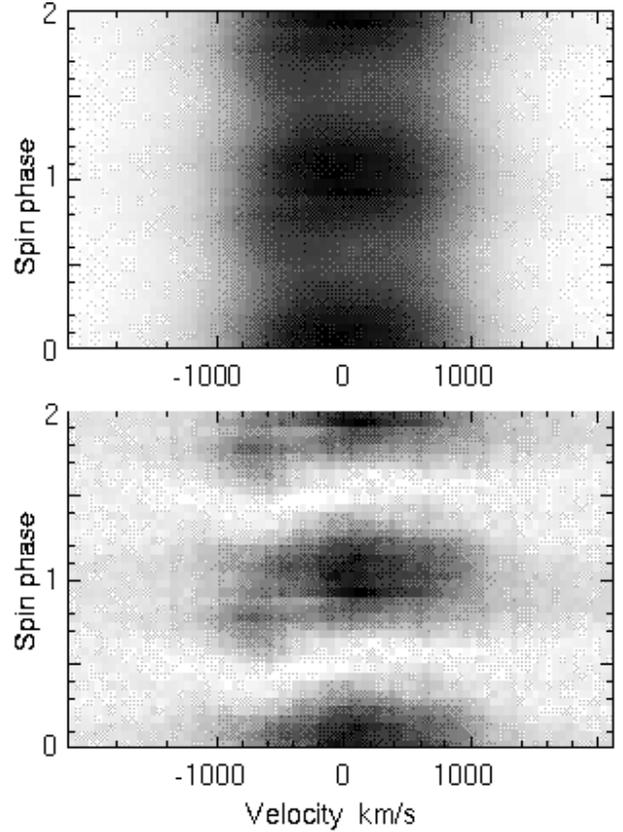}
\caption{The upper panel shows the \Hb\ line folded on the spin period,
while the lower panel shows the same after subtraction of the phase-invariant
profile. For future reference, the adopted phase zero corresponds to 
HJD 2450213.88813.}
\end{figure}

The timescale for
spinning up the white dwarf to equilibrium is 10$^{6}$--10$^{7}$ yrs
(assuming an accretion rate of 10$^{16}$ g s\minone, typical of systems 
below the period gap), which is much 
shorter than the \sqiggt 10$^{8}$-yr evolutionary timescale. Thus we would
require an explanation for EX~Hya and \vcen\ being far
from equilibrium. The most plausible explanation is a change in the
mass-transfer rate.  For instance, if mass transfer shut off for a
long period, the spin period could lengthen and become locked to the
orbit, in the manner of an AM~Her star; a resumption of mass transfer
would then spin up the white dwarf again, heading for \Pspin \sqig 0.1
\Porb.  Note that the white dwarf
in EX~Hya is currently spinning up on a timescale of 4\pten{6}\,yrs
(e.g.\ Jablonski \&\ Busko 1985), in line with this picture, although
with only a 30-yr span of observations we can't distinguish between a
sustained spin-up and a short-term fluctuation.

\subsection{The line profiles and tomogram}
The orbital tomogram shows that the high-velocity emission is not
symmetric about the white dwarf, being enhanced in the lower-left
quadrant. Thus the higher-velocity regions (the inner
disc or magnetosphere) are not symmetric over orbital phase. Since \vcen\
is not eclipsing, it is unlikely that this is due to obscuration of
the inner regions by disc structure.  Thus the only likely possibility
is that the accretion stream overflows the outer disc and continues
into the inner disc, where it creates a disturbance localised in
orbital phase. This has previously been proposed in IPs to explain
X-ray beat periods (e.g.\ Hellier 1991) and might be occurring in
non-magnetic cataclysmic variables such as SW~Sex stars (e.g.\ Hellier
\& Robinson 1994). The idea also has theoretical support (e.g.\ Armitage
\&\ Livio 1996; 1998). 

In the SW~Sex stars the high-velocity line wings are at maximum
redshift at orbital phase \appro 0.9 (e.g.\ Thorstensen \etal\ 1991; 
Hellier 1996), which
matches that in \vcen\ with our adopted phasing. The brightest region
in the \vcen\ tomogram is then at the right phase to correspond to
emission from either the secondary star or the bright spot where the
stream hits the disc (or a mixture of these; the phase uncertainty
prohibits a secure distinction between the possibilities). Weaker
emission is then seen looping leftwards towards the higher-velocity
feature, and could be emission from the overflowing stream.

We don't have sufficient information (masses and inclination) to
interpret the velocities in the tomogram directly, but we can perform
a plausibility check. The high-velocity wings in eclipsing 
SW~Sex stars extend to velocities of \appro 1400--1600 \kmps\
in stars such as SW Sex itself (Dhillon, Marsh \&\ Jones 1997) and
V1315 Aql (Hellier 1996); 
the equivalent component in \vcen\ extends to \appro 1300 \kmps\ 
in data with a comparable signal-to-noise ratio. These velocities match if
\vcen\ has an inclination of \appro 60\deg, or a $\sin i$ of \appro 0.87.

Further, adopting a white-dwarf mass of 0.7 \msun\ and a 
red-dwarf mass of 0.1 \msun\ implies that the red dwarf has an orbital
velocity of \appro 440 \kmps, that the Lagrangian point orbits at \appro
290 \kmps, and that the outer edge of the disc (assuming it is located at the
tidal limit) orbits at \appro 650 \kmps\ (see Warner 1995, chapter 2, 
for the relevant formulae).  These values compare with the
observed S-wave amplitude of 350 \kmps, or \appro 400 \kmps\ with the
above $\sin i$. Thus the S-wave is compatible with arising from the 
secondary or the early part of the stream, but less compatible with arising
from the stream--disc impact (unless the inclination or the white-dwarf
mass are lower than adopted above).  Thus, overall, the
line profiles are consistent with the stream-overflow idea, in that
both the lower-velocity S-wave and the line wings have compatible 
velocities.

\begin{figure*} \vspace{6.5cm}      % Fig 5
\includegraphics{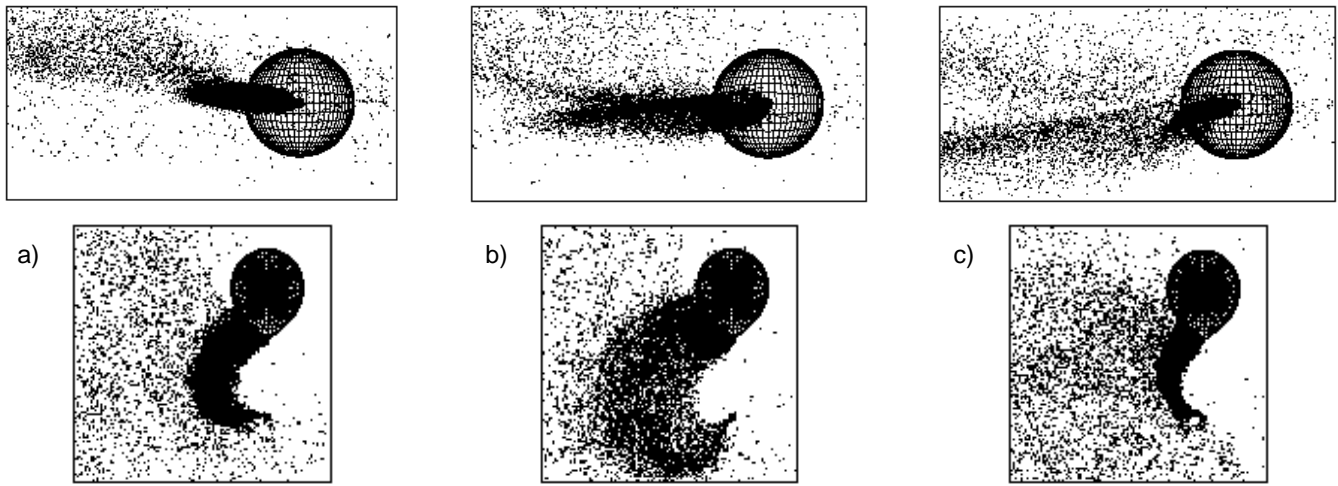}
\caption{Illustrations of the flow in the diamagnetic-blob model
at different beat-cycle phases (a to c).  In each case we
show both a view along the line of stellar centers (rectangular panels) 
and a plan view (squares). Note the change from material being pushed 
above the plane (panel a) to being pushed below the plane (panel c) as 
the dipole orientation changes. The phases are 0.12 (a), 0.30 (b) and 
0.49 (c), defining beat-cycle zero as occuring
when the upper magnetic pole points towards the secondary.}
\end{figure*}

\subsection{The periodicities}
A crucial observation for models of \vcen\ is that the X-ray
lightcurve varies only with the 2147-s spin period, and not with the
orbital cycle nor the orbital sidebands of the spin period (Hellier
\etal\ 1998).  This implies that the accreting material loses
knowledge of orbital phase before attaching to field
lines. This, in turn, suggests that, if the stream-overflow model is
correct, the overflowing stream does not travel far enough to
encounter the magnetosphere, but instead re-impacts the disc further
out.  This contrasts with suggestions for other IPs, for instance
FO~Aqr, where the interaction of the overflowing stream with the
magnetopshere was invoked specifically to explain an X-ray beat pulse
(Hellier 1993; Beardmore \etal\ 1998). Two caveats should be made.  First, the overflow might
be intermittent, and might not have been occurring during the X-ray
observations. Indeed, the X-ray beat pulse in FO~Aqr is variable and
sometimes absent. Secondly, we should consider whether 
the 2147-s period is misidentified, and is actually the beat (\beat)
period. This, though, would imply a spin period of 1509 s, and no such
periodicity has ever been seen in \vcen; and further, other observed
periodicities, such as the 1860-s modulation, would then have no
natural identification.

There is, however, a significant beat-cycle modulation in the 
line V/R ratios (Section~2). This can be explained in the standard way
for optical beat periods, namely irradiation of structure fixed in the
binary frame (secondary or stream) by the spin-pulsed X-ray beam. 
However, the period of the observed modulation differs from the 
expected value by 0.7 per cent. Over the 2-d span of the observations 
this amounts to a shift of 0.35 cycles. A possible explanation is that over
the 2-d interval the X-ray beam switched from illuminating (predominantly)
the secondary, to illuminating (predominantly) the structure formed where
the overflowing stream re-impacts the disc. As can be seen from the tomogram,
these two regions are separated by \sqig 0.35 in orbital phase. If correct,
this again sugggests that the overflow is intermittent, occuring only some
of the time. 

One puzzle for the above model is the observation of the 1860-s photometric
modulation, identified with 2(\beat), when \beat\ is not seen. 
Reprocessing of X-rays would likely result in an optical \beat\ 
modulation, as is observed in many IPs, but not 2(\beat). One 
plausible explanation, that the illuminating X-ray beam is double-peaked, 
resulting in reprocessing at 2(\beat), is contradicted by the fact 
that the observed X-ray pulse is nearly sinusoidal 
(Hellier \etal\ 1998). Thus, this explanation only works if the X-ray 
pulse is beamed such that it is double-peaked in the orbital plane
but sinusoidal from our line of sight, which is unlikely.

\section{A discless model?}
Having considered a model for \vcen\ based on the conventional partial
disc, we now consider the alternative discless model based on a 
diamagnetic flow. This
model was proposed by King (1993) and Wynn \&\ King (1995), see also
Wynn (2001).  It treats the accretion flow as a set of diamagnetic
blobs, and represents these by the particles in a hydrodynamical code,
with the addition of a magnetic drag term which acts like the tension 
of the magnetic field lines.
This term is proportional to the rate at which particles cross
field lines, giving an acceleration \[ \mbold{a}_{\rm mag} = -k
[\mbold{v} - \mbold{v}_{\rm f}]_{\perp}\] where $\mbold{v}$ and
$\mbold{v}_{\rm f}$ are the velocities of the material and field and
the symbol $\perp$ indicates the component perpendicular to the field
lines. The parameter $k$ is dependent on factors such as the the local
field strength, blob density and Alfv\'en speed.  The net effect is
that particles orbiting outside the corotation radius gain angular
momentum from the field, and can be pushed outwards, while particles
orbiting inside the corotation radius lose angular momentum to the
field and so accrete onto the white dwarf.

King \&\ Wynn (1999) proposed that this model can explain the 
anomalously long spin periods of EX~Hya and \vcen, suggesting that the 
systems are in an equilibrium where \rco\ $\approx$ $b$. Using this model
we have computed simulations of the accretion flow appropriate to \vcen.
We assume white-dwarf and red-dwarf masses of 0.7 and 0.1 \msun\ 
respectively. Using an orbital period of 5077 s we then tweak the
$k$ parameter until we get a spin period of 2147 s.   This requires
a magnetic timescale, $k^{-1}$, of a few seconds.  We can combine this
with estimates for the density and blob-length in the stream [10$^{-9}$
g cm$^{-3}$ and 10$^{9}$ respectively, see King \&\ Wynn (1999)] to
find a magnetic moment of \appro 5\pten{32}\  G cm$^{3}$ (equating to 
a field of $\approx$ 1 MG).

In this model the flow alternates between episodes of accretion and
ejection, according to the beat phase between the orbital cycle and
the white-dwarf rotation.  Accretion events occur when one of the
magnetic poles points towards the accretion flow, 
allowing the blobs to flow down field lines; they thus
occur twice per beat cycle. Between each accretion event, when the
magnetic poles are on the white-dwarf limb as seen from the
approaching flow, the blobs are expelled outwards, and may be swept up
by the secondary.

Most of the flow (\sqiggt 90 per cent) accretes, but the expulsion of the 
remaining \sqig 10 per cent, with a high specific angular momentum, allows the
system to maintain equilibrium at a far longer spin period than would
be possible in a disc-fed system.  Illustrations of the flow in this
model are presented in Fig.~5. 

\begin{figure} \vspace{14.7cm}      % Fig 6
\includegraphics{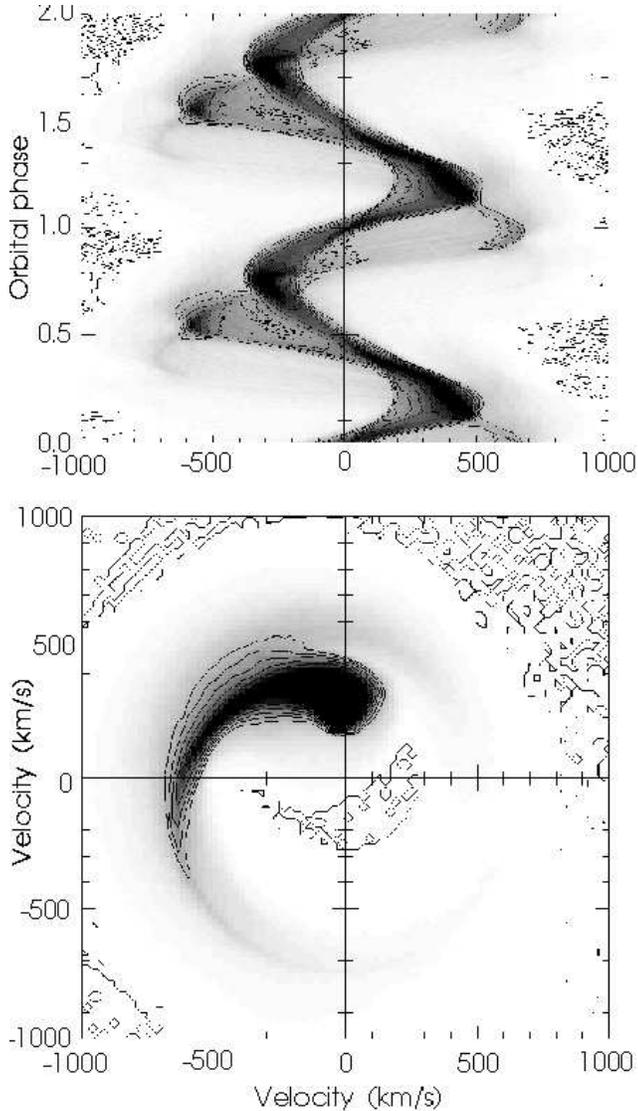}
\caption{The trailed spectrum from the diamagnetic-blob model together
with the corresponding tomogram. The secondary is at inferior conjunction
at phase zero.}
\end{figure}

\subsection{Sideband pulsations in the diamagnetic-flow model}
Perhaps the biggest difficulty in applying this model to EX~Hya and
\vcen\ is that it predicts an accretion rate modulated on the beat
cycle (e.g.\ fig.~3 of King \&\ Wynn 1999), whereas the hard-X-ray
lightcurves of both stars are modulated only at the spin frequencies
(e.g.\ C\'ordova, Mason \&\ Kahn 1985; Hellier \etal\ 1998). The
current models calculate only the blob-accretion rate and don't yet
predict X-ray lightcurves --- this would involve factors such as the
optical depth in the accreting regions, which could modulate the flux
at the spin frequency. However, the lightcurve would still be expected
to show the hallmarks of discless accretion, namely pulsations at
$\Omega$, \beat\ and/or $2\omega$\,--\,$\Omega$ (e.g.~Wynn \&\ King 1992),
whereas none of these are seen in the $>$\,2 \kev\ X-ray lightcurves
of either star. This argument is supported by V2400~Oph,
an IP which is almost certainly discless, whose X-ray lightcurve
is dominated by the beat pulse (Buckley \etal\ 1997; Hellier \&\
Beardmore 2002). 

Having said the above, the simple magnetic-drag prescription adopted in
the current model may not be appropriate close to the white dwarf, since it 
ignores magnetic pressure 
and Kelvin--Helmholtz instabilities. Also, the 
equilibria of EX~Hya and \vcen\ require only \sqig 10 per cent of the flow
to be ejected back towards the secondary. It is thus conceivable
that the bulk of the flow circularizes into a azimuthally symmetric,
structure around the white dwarf, which could then result in
a dominant spin-cycle pulsation.  In particular, if there were a range
of blob densities, the denser blobs would be less affected by the 
field, and circularise into a disc more easily. The equilibrium could
then be sustained by the ejection of less-dense blobs.

Optical beat-cycle pulsations are less diagnostic, since they can be
created by other mechanisms, including the irradiation of structure
fixed in the orbital frame by spin-pulsed X-rays. As discussed in 
Section~5.2, this is likely to explain the optical beat-frequency 
pulsation seen
in \vcen, and applies equally well to the diamagnetic-blob model. Again, 
the 0.7 per cent difference in the observed frequency from the true 
frequency requires that the illuminated structure 
changes location from day to day --- and might be expected as the 
stream is being buffeted by a varying magnetic force.
On the other hand, the detection of the 1860-s optical pulsation
identified with 2(\beat) strongly supports
the diamagnetic-blob model.  The key feature of this model, ejection
events twice per beat cycle, offers a straightforward explanation of
an optical 2(\beat) periodicity.

\subsection{Tomographic comparison}
The diamagnetic-blob model can be used to predict line profiles simply
by adding up the number of blobs in each velocity bin. Although this
ignores all radiative transfer effects, it is still a useful comparison
with the data. Fig.~6 shows the trailed spectra from the model, 
computed for an inclination of 90\deg, along
with an orbital-cycle tomogram.  The trailed spectrum shows similarities
with the data in Fig.~3, in that both have a lower-velocity S-wave 
accompanied by higher-velocity emission phased 0.2--0.3 earlier.

In the tomogram both the data and the model show a `hook' structure,
which is brightest near the secondary, and curls 
anti-clockwise before petering out in the lower-right quadrant. Note
that the velocities of the hook feature in the model are lower
than expected for disc emission, since they arise from material being
ejected back towards the secondary.  If \vcen\ is at a relatively high
inclination, so that the $\sin i$ factor is \appro 1, then the model
and observed velocities match well.

\begin{figure*} \vspace{11.8cm}      % Fig 7
\includegraphics{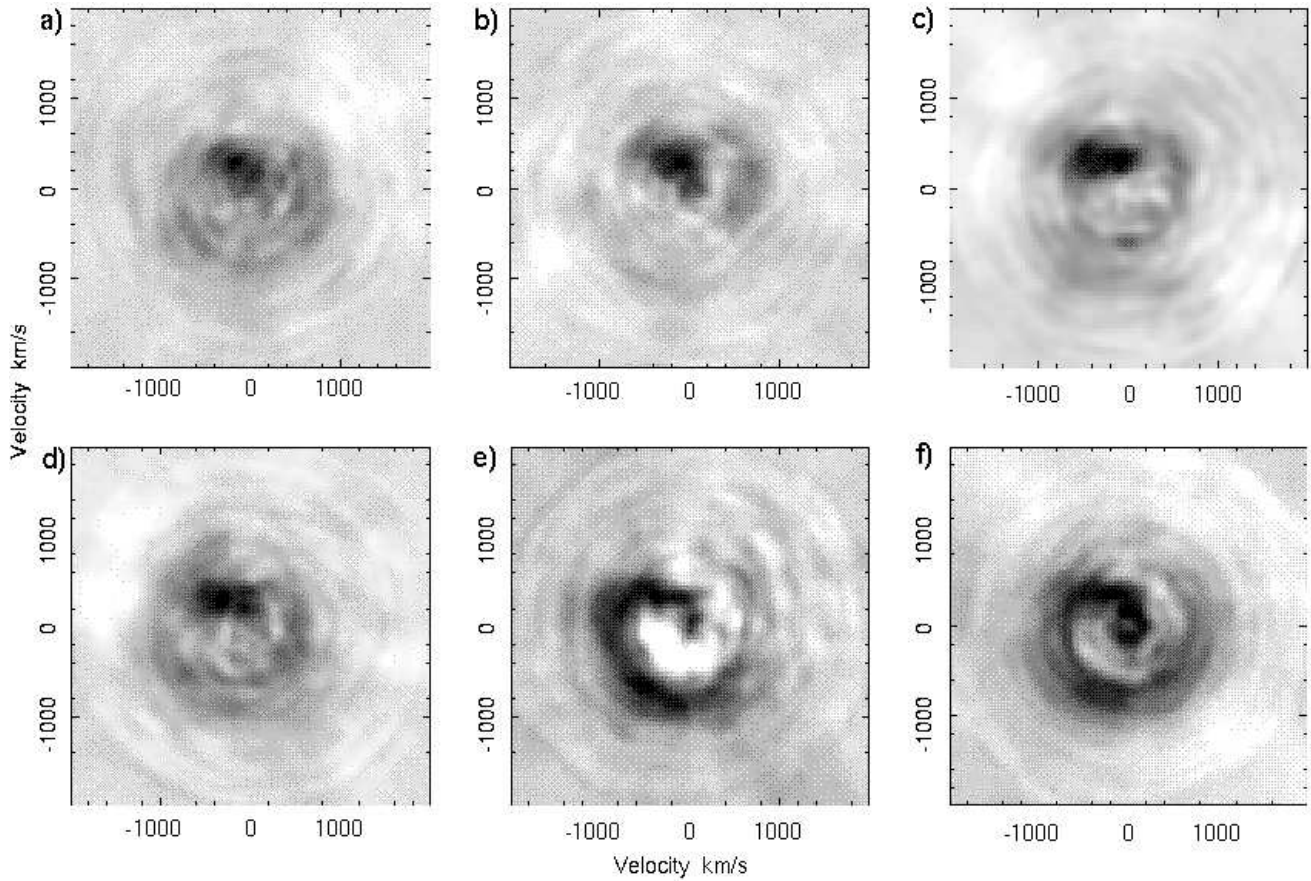}
\caption{Orbital-cycle tomograms of the \Hb\ line from selected
ranges of beat-cycle phase. The ranges are (panels a to f) 0.0--0.19, 
0.19--0.38, 0.38--0.56, 0.56--0.75, 0.75--0.94 and 0.94--1.13.
We define beat phase as zero when the upper pole points towards the
secondary, as in Figs.~5 \& 8. However, when dealing with the data, this
is dependent on the correctness of our interpretations of Fig.~3 (inferior
conjunction of the secondary at phase 0) and of Fig.~4 (upper pole pointing
away at phase 0) which are both insecure.}
\end{figure*}

This similarity in tomograms is, along with the explanation for the 
long spin period, the strongest evidence for a discless flow in \vcen.
However, as discussed above, a discless flow is dependent primarily
on the beat cycle, rather than the orbital or spin cycles. Thus we can
go further and compare orbital-cycle tomograms that are sampled
from particular beat phases, an analysis technique that has been used
previously for FO~Aqr data (Marsh \&\ Duck 1996).

\begin{figure*} \vspace{11.8cm}      % Fig 8
\includegraphics{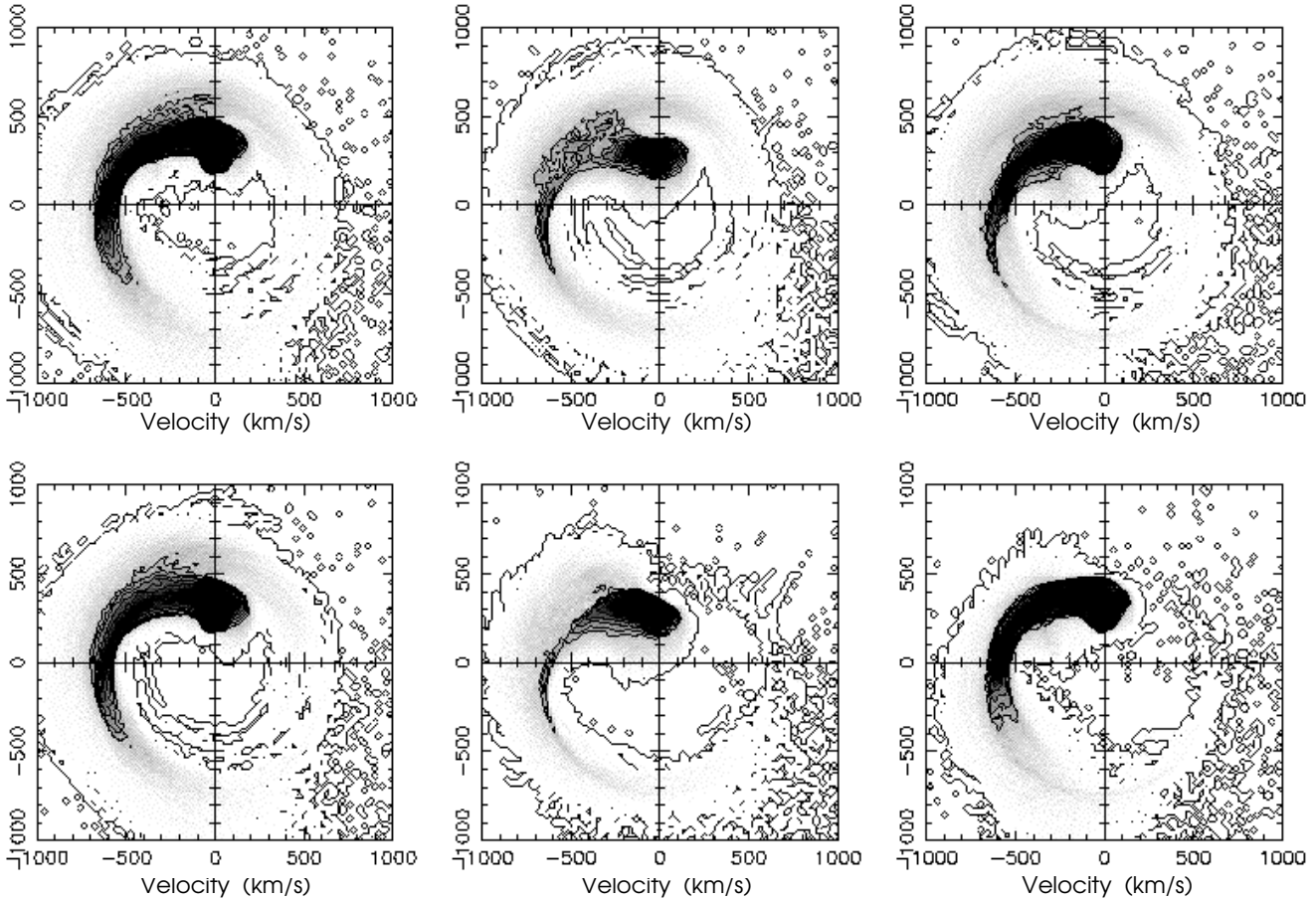}
\caption{Model orbital-cycle tomograms from selected 
ranges of beat-cycle phase. The sequence and ranges are the same as in
Fig.~7. Beat-cycle zero occurs when the upper magnetic pole points
towards the secondary.}
\end{figure*}

Before proceeding we should express a strong caveat about any such
method.  Given a particular orbital and beat phase, the spin phase is
entailed.  Thus if there are variations that are intrinsic to the spin
cycle, from magnetically threaded material near the white
dwarf, then these will not smear out in the analysis, but will be
systematically folded in, appearing as structure in the orbital
tomogram, varying according to the beat phase of the sample. 

Thus the
beat-phase-resolved orbital tomogram will have superimposed on it the
tomogram of the spin cycle (see Hellier 1999 for spin-cycle tomograms
of IPs) and this superimposed, spin-cycle tomogram will rotate
anti-clockwise as the beat phase of the sample increases.  Thus we
expect to find changes in the orbital tomogram as a function of beat
phase, even if there are no intrinsic variations on the beat cycle!
In the case of \vcen, the change over the spin cycle is mainly in
equivalent width, rather than velocity, and this will reduce the
contamination.  In the case of FO~Aqr, the lines vary over the
spin cycle in both equivalent width and velocity (e.g.\ Hellier, 
Cropper \&\ Mason 1990) and the contamination will be severe. 
More generally, there is no model-independent way of
attributing variations to the beat cycle rather than the spin cycle,
once these are allowed to be functions of orbital phase.

A further caveat is a standard one applying to tomography, in that it
assumes that optical depth effects are not changing the intensity of
line emission over the different cycles.  From the large changes in
equivalent width, particularly over the spin cycle (Fig.~2), we know
that this assumption is violated in \vcen.

In Fig.~7 we present beat-phase-resolved, orbital-cycle tomograms of
the \vcen\ data.  The changes in features over beat phase will be
caused by (1) components whose velocities genuinely vary with beat
phase, (2) components whose velocities vary with spin phase, and (3)
material whose illumination by EUV and X-ray photons is a function of
beat phase.  For comparison, Fig.~8 shows the equivalent tomograms
computed from the diamagnetic-blob model. These will contain the
effects 1 and 2 just mentioned, but not effect 3, since irradiation is
not included.

Both the data and the model tomograms show the hook-like features 
varying with beat phase.  In the model, the `tail' of the hook  
becomes less pronounced during the ejection events, re-forming during
accretion episodes. This disappearance and re-forming 
is also seen clearly in the data. However, the variation is greater 
in the data, and occurs once per beat cycle, whereas in the model it 
occurs twice per beat cycle. The difference can plausibly be explained
by the lack of X-ray irradiation in the model. Given the sinusoidal
X-ray spin pulse, regions fixed in the orbital frame will receive a 
sinusoidal cycle of irradiation over the beat cycle, 
enhancing the line emission once per beat cycle.  Given that
irradiation is not included in the model, and the other caveats expressed
above, it is not surprising that there are major differences between the
data and the current model.

To complete the discussion of the beat-resolved tomograms (Fig.~7) we 
should consider how they could be interpreted in the partial-disc model. 
The emission near velocity (0, 300) \kmps, seen at all beat phases,
would likely be from the secondary star. The `hook' feature would
be due to the stream flowing from the Lagrangian point, colliding with
the disc (near velocity --500, 400 \kmps), and then overflowing the
disc (moving to the lower-left quadrant of the tomogram). The fact
that the velocities in the lower-left quadrant are only about half
those of a free-falling stream means that the overflowing stream
would have to be slowed by its interaction with the disc, though this
is in line with theoretical findings (Armitage \&\ Livio 1998).

The stream would be illuminated once per beat cycle by the spin-pulsed
X-ray beam, and this would explain why it is seen only for 
particular beat phases.  However, we don't have an ephemeris for the
X-ray pulse and so cannot reliably predict which phases these are. 

\section{Conclusions}
We have presented phase-resolved spectroscopy of the
intermediate polar \vcen. We summarise here the strengths and weaknesses
of the two models proposed for this system. 

\subsection{Partial-disc model}
The strengths of the model are: (1) 
Disc-fed accretion explains an X-ray lightcurve
varying only at the spin period. (2) 
The line-profile variations can be plausibly explained
by invoking stream--disc overflow, in a manner seen in SW~Sex stars
(although a weak X-ray beat pulsation might then be expected). (3) 
The beat-resolved tomograms can be interpreted as showing
structure from the stream overflowing the disc, illuminated once
per beat cycle by the X-ray beam. 

The weaknesses are: (1) 
There is no explanation for the anomalously long spin
period of \vcen, except to claim that the system is not in equilibrium.
(2) There is no easy explanation for the 1860-s optical pulsation.

\subsection{Discless model}
The strengths of the model are: (1) 
Explains the long spin period of \vcen. (2) 
Explains the hook-like features in the tomograms, and the
changes in the feature over beat-cycle phase (though there are still
differences with the data). (3) Explains the 1860-s optical pulsation 
as ejection events occurring at 2(\beat).  The main weakness is the 
fact that the X-ray lightcurve varies only at the spin
frequency, and not at the orbital and beat frequencies, which 
argues against stream-fed accretion. 

\subsection{Further work}
The summary above shows that we can't yet decide between the
two models.  Note that there is also a scenario in which both may be 
`right': if \vcen\ had been discless for most of the past 
10$^{6}$--10$^{7}$ y, this would explain the long spin period, even if
a disc had formed more recently.  

Observations needed to make further progress include a
determination of the inclination of \vcen, allowing us to match the
observed line velocities to the models. Developments to the theory
could include the addition of radiation processes to the
diamagnetic-blob model, allowing a better comparison with line profiles,
and computations of the predicted X-ray lightcurves. Also useful would
be deeper searches for polarisation, following the initial work by
Buckley \etal\ (1998).  In the diamagnetic-flow model the field
is likely to be an order-of-magnitude stronger than in a disc-fed
system of the same orbital period (fig 7 of King \&\ Wynn 1999), and
further the polarised light would not be diluted by
a bright disc, leading to a greater likelihood of detecting
polarisation in these systems than in most IPs.


\begin{thebibliography}{}
\bibitem[]{}Allan A., Hellier C., Buckley D.\,A.\,H., Beardmore, A.\,P.,
     1999, in Hellier C.,
     Mukai K., eds, Annapolis Workshop on Magnetic Cataclysmic Variables, 
     ASP Conf.~Ser., 157, p57
\bibitem[]{}Armitage P.\,J., Livio M., 1996, ApJ, 470, 1024
\bibitem[]{}Armitage P.\,J., Livio M., 1998, ApJ, 493, 898
\bibitem[]{}Beardmore A.\,P., Mukai K., Norton A.\,J., Osborne J.\,P.,
      Hellier C., 1998, MNRAS, 297, 337
\bibitem[]{}Buckley D.\,A.\,H., Cropper M.\, Ramsay G., 
        Wickramasinghe D.\,T., 1998, MNRAS, 299, 83
\bibitem[]{}Buckley D.\,A.\,H., Haberl F., Motch C., Pollard K.,
        Schwarzen\-berg\--Czerny A., Sekiguchi K., 1997, MNRAS, 287, 117
\bibitem[]{}C\'ordova F.\,A. Mason K.\,O., Kahn S.\,M., 1985, MNRAS, 212, 447
\bibitem[]{}Dhillon V.\,S., Marsh T.\,R., Jones D.\,H.\,P., 1997, MNRAS, 
        291, 694
\bibitem[]{}Hellier C., 1991, MNRAS, 251, 693
\bibitem[]{}Hellier C., 1993, MNRAS, 265, L35
\bibitem[]{}Hellier C., 1996, ApJ, 471, 949
\bibitem[]{}Hellier C., 1999, ApJ, 519, 324
\bibitem[]{}Hellier C., 2001, Cataclysmic Variable Stars, Springer--Verlag,
        Heidelberg
\bibitem[]{}Hellier C., Beardmore A.\,P., 2002, MNRAS, in press
\bibitem[]{}Hellier C., Beardmore A.\,P., Buckley D.\,A.\,H.,  1998,
        MNRAS, 299, 851
\bibitem[]{}Hellier C., Mason K.\,O., Cropper M., 1990, MNRAS, 242, 250
\bibitem[]{}Hellier C., Mason K.\,O., Rosen S.\,R., Cordova F.\,A.,  
               1987, MNRAS, 228, 463
\bibitem[]{}Hellier C., Robinson E.\,L., 1994, ApJ, 431, L107
\bibitem[]{}Jablonski F., Busko I.\,C., 1985, MNRAS, 214, 219
\bibitem[]{}King A.\,R., 1993, MNRAS, 261, 144
\bibitem[]{}King A.\,R., Wynn G.\,A.,  1999, MNRAS, 310, 203
\bibitem[]{}Marsh T.\,R., Duck S.\,R., 1996, NewA, 1, 97
\bibitem[]{}Marsh T.R., Horne K., 1988, MNRAS, 235, 269
\bibitem[]{}Thorstensen J.\,R., Ringwald F.\,A., Wade R.\,A., Schmidt G.\,D.,
    Norsworthy J.\,E., 1991, AJ, 102, 272
\bibitem[]{}Warner B., 1995, Cataclysmic Variable Stars, Cambridge University
        Press, Cambridge
\bibitem[]{}Wynn G.\,A., 2001, in Boffin H., Steeghs D., Cuypers J., eds,
                    Astrotomography, Springer--Verlag, Heidelberg, p155
\bibitem[]{}Wynn G.\,A., King A.\,R., 1992, MNRAS, 255, 83
\bibitem[]{}Wynn G.\,A., King A.\,R., 1995, MNRAS, 275, 9
\bibitem[]{}Wynn G.\,A., King A.\,R., Horne K., 1997, MNRAS, 286, 436
\end{thebibliography}
\end{document}